\documentclass[twoside]{dis04}
\def\runauthor{Raimund Str\"ohmer}
\def\shorttitle{W-mass and the WW and ZZ crossection at LEP}
\newcommand {\W} {\ensuremath{\mathrm{W}}}
\newcommand {\Z} {\ensuremath{\mathrm{Z}}}
\newcommand {\Znull} {\ensuremath{\mathrm{Z}^0}}
\newcommand{\TEVATRON}{\ensuremath{\mathrm{\hbox{Te\kern -0.1em Vatron}}}}
\newcommand{\LEP}{\ensuremath{\mathrm{\hbox{LEP}}}}

\newcommand {\GeV}    {\ensuremath{\mathrm{Ge\kern -0.1em V}}}
\newcommand {\MeV}    {\ensuremath{\mathrm{Me\kern -0.1em V}}}
\newcommand{\qqqq}    {\mathrm{q\bar{q}q\bar{q}}}
\newcommand{\qqln}    {\mathrm{q\bar{q}l\nu}}
\newcommand{\PLB}[3]  {Phys.\ Lett.\ {\bf B#1} (#2) #3}
\newcommand{\NPB}[3]  {Nucl.\ Phys.\ {\bf B#1} (#2) #3}
\newcommand{\PRB}[3] {Phys.\ Rev.\ {\bf B#1} (#2) #3}
\newcommand{\PRD}[3]  {Phys.\ Rev.\ {\bf D#1} (#2) #3}
\newcommand{\CPC}[3]  {Comp.\ Phys.\ Comm.\ {\bf #1} (#2) #3}
\begin{document}

\title{Determination of the W-mass and the WW \\
and ZZ cross-section at LEP}

\author{Raimund Str\"ohmer}

\address{LMU M\"unchen, Sektion Physik\\
Am Coulombwall 1, D-85748 Garching, Germany\\
E-mail: Raimund.Stroehmer@Physik.Uni-Muenchen.DE }

\maketitle

\abstracts{
We review the precision measurement of the mass 
of the W-boson at LEP. 
We discuss the techniques used by the four LEP experiments
to determine the mass of the W-boson as well as the major sources of 
systematic uncertainty.
The measurement of the $\W^+\W^-$~and ZZ cross-sections are presented.  
}

\section{Introduction}
In 1996  the center-of-mass energy of the electron-positron collider 
LEP was increased  above the
threshold of two times the W-boson mass ($M_\W$) which made it possible 
to  produce pairs of W-bosons in 
$e^+e^-$ collisions
and thus opened the opportunity for a precision determination of the
\W-boson mass and measurements of its couplings and decay 
branching ratios.
A detailed review of measurements of the W-boson mass and its couplings
can be found e.g. in~\cite{bib:rst-paper}.

\section{Motivation}
In the electroweak Standard Model, the properties of the \Znull- and W-boson depend 
on a few fundamental parameters  only.
The comparison of the directly measured W-boson mass with Standard Model
predictions based on precision measurements of \Znull-boson properties is 
therefore an important test of the Standard Model.
In the lowest order calculation (at tree level),
$M_\W$ only depends on the Fermi constant $G_\mu$, which is accurately known from
muon decays, the fine structure constant $\alpha$ and the mass of the \Znull-boson $M_Z$.
Loop corrections lead
to a quadratic dependence of $M_\W^2$ on the top mass and a logarithmic dependence
on the Higgs mass.
\begin{figure}[!thb]
\parbox{5.6cm}{
\epsfig{file=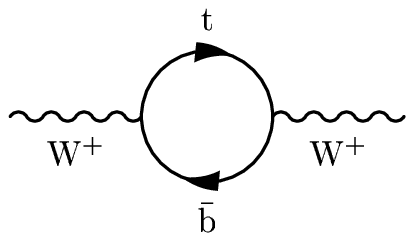,width=5.5cm} 
\epsfig{file=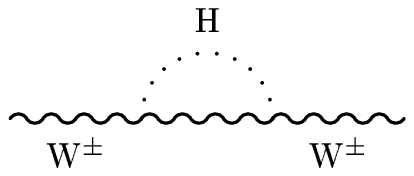,width=5.5cm}
\caption{Feynman diagram of 1-loop corrections to the \W~propagator.
 \label{fig:mw_loop}}
}
\hspace*{.3cm}
\parbox{6.2cm}{
\epsfig{file=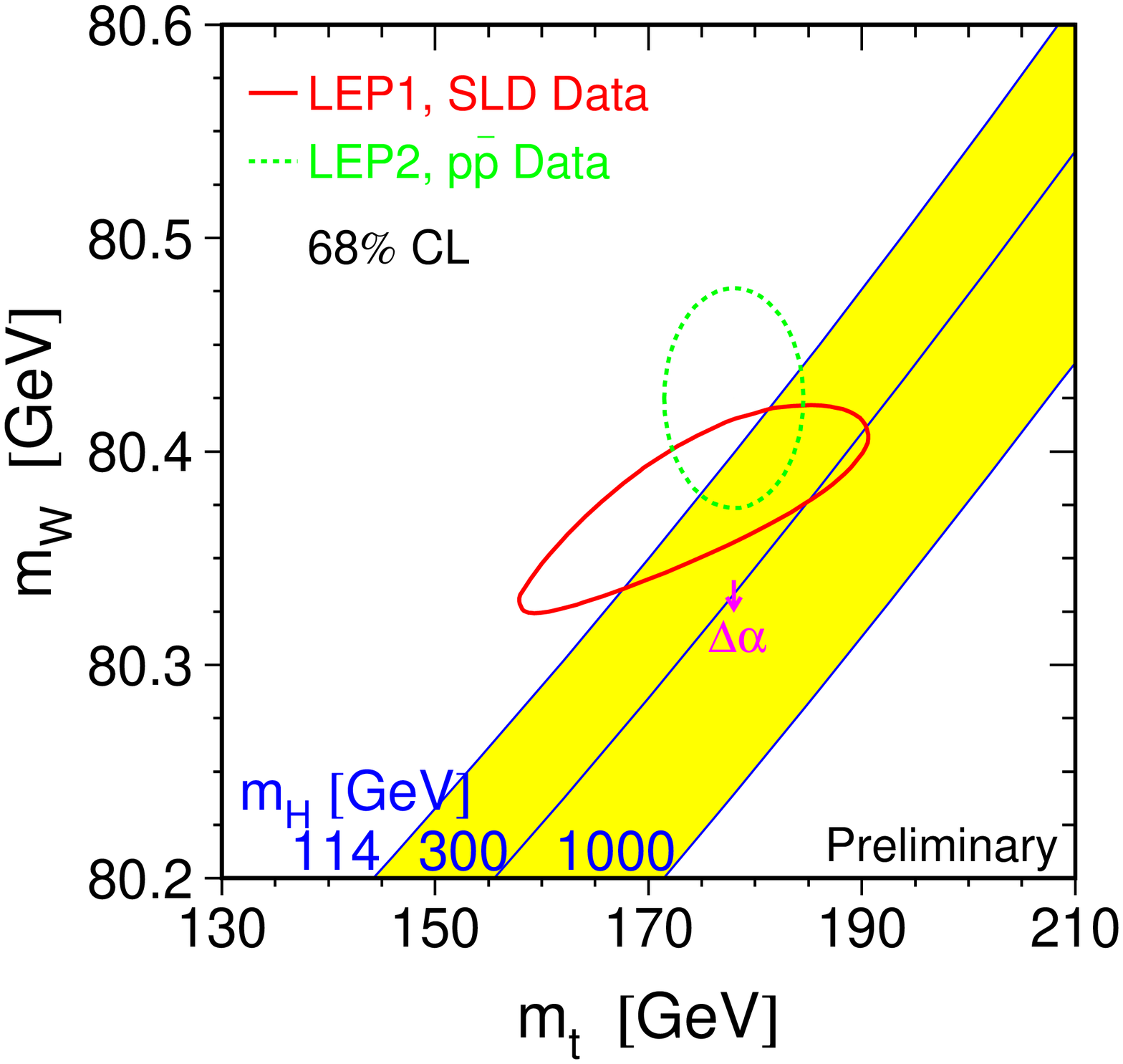,width=6.1cm}
\caption{Result of a Standard Model fit to the electroweak data
and the Standard Model prediction of the \W-boson mass as 
function of the top mass.
\label{fig:mw_mt} }
}
\end{figure}
As an example, Figure  \ref{fig:mw_loop} shows 1-loop
contributions to the W propagator including a top quark and a Higgs boson.
Figure \ref{fig:mw_mt} shows the prediction for the \W-boson
and top quark mass from a fit to all data excluding the direct measurements of $M_\W$~and $M_{top}$~\cite{bib:elw}.
The figure also shows the Standard Model prediction for $M_\W$~as function of $M_{top}$
for three  different Higgs masses. The predictions are compared to
direct measurements of the top quark mass
and the W-boson mass at the \LEP~and Tevatron colliders. 
This comparison  is an important test of
the electroweak Standard Model.

The measurement of \W-pair and other  four-fermion cross sections
can be used to study triple gauge boson couplings.
Any deviations from the Standard Model predictions can be interpreted
as a sign for physics beyond the Standard Model.

\section{W-Pair Production and Decay}
\begin{figure}[!thb]
\begin{center}
\epsfig{file=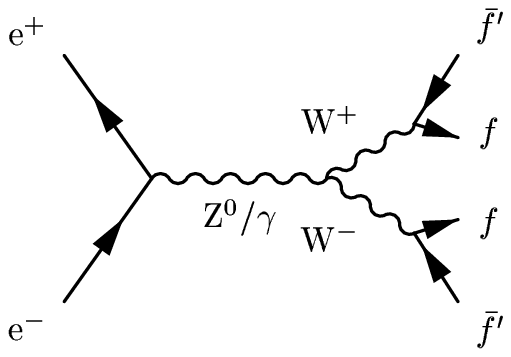,width=4.cm}
\epsfig{file=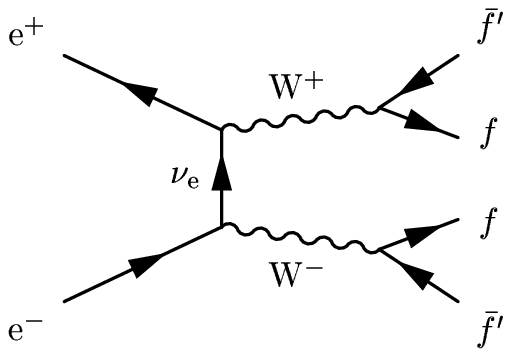,width=4.cm}
\end{center}
\caption{Feynman diagrams for \W-pair production.
 \label{fig:fey_ww}}
\end{figure}
Above the center-of-mass energy threshold of $  2 \cdot M_\W$,  \W-boson pairs can 
be produced in $e^+e^-$ annihilation.
Figure \ref{fig:fey_ww} shows the tree level Feynman diagrams contributing
to the \W-pair production (called CC03). 
 The Feynman diagrams for the \W-pair production via a 
virtual Z or photon contain a triple gauge boson coupling which can therefore
be studied by measuring the pair production cross section.

The W-boson decays in 68.5\% hadronically into a quark-antiquark pair,
which will be observed in the detector as two jets, and in 31.5\% leptonically
into a charged lepton and a neutrino.
Depending on the decay of the two W-bosons this leads to three
distinct signatures. In 46\% of the events, both W-bosons 
decay hadronically (hadronic decays). For these events, one expects four 
jets.
In 44\% of the events, one W decays hadronically and the other leptonically
(semileptonic decay), leading to events with two jets, a high energetic
lepton and missing energy due to the unobserved neutrino. 
For the 10\% of events in which both W-bosons decay leptonically, the event 
only contains two high energetic leptons and a large amount of missing 
energy. 
In order to study the different W-boson pair decays, it is important to be able to
precisely measure the momentum and direction of the leptons and jets.

\section{W- and Z- Pair cross-section}

\begin{figure}[!thb]
\parbox{6.0cm}{
\epsfig{file=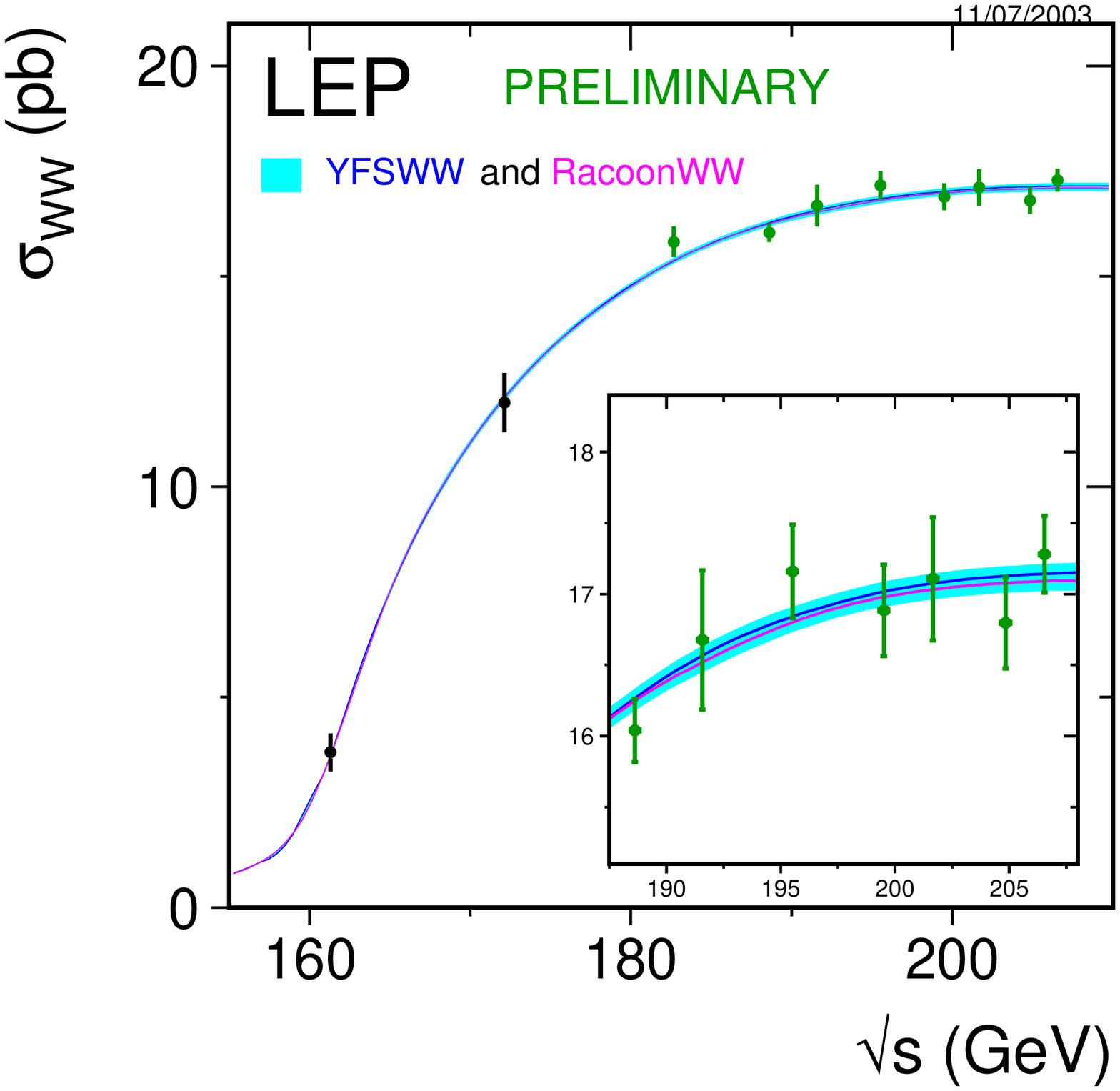,width=6.0cm}
\caption{ W-pair cross-section as function of center-of-mass energy.
 \label{fig:w_crossa}}
}
\hspace*{.3cm}
\parbox{6.0cm}{
\epsfig{file=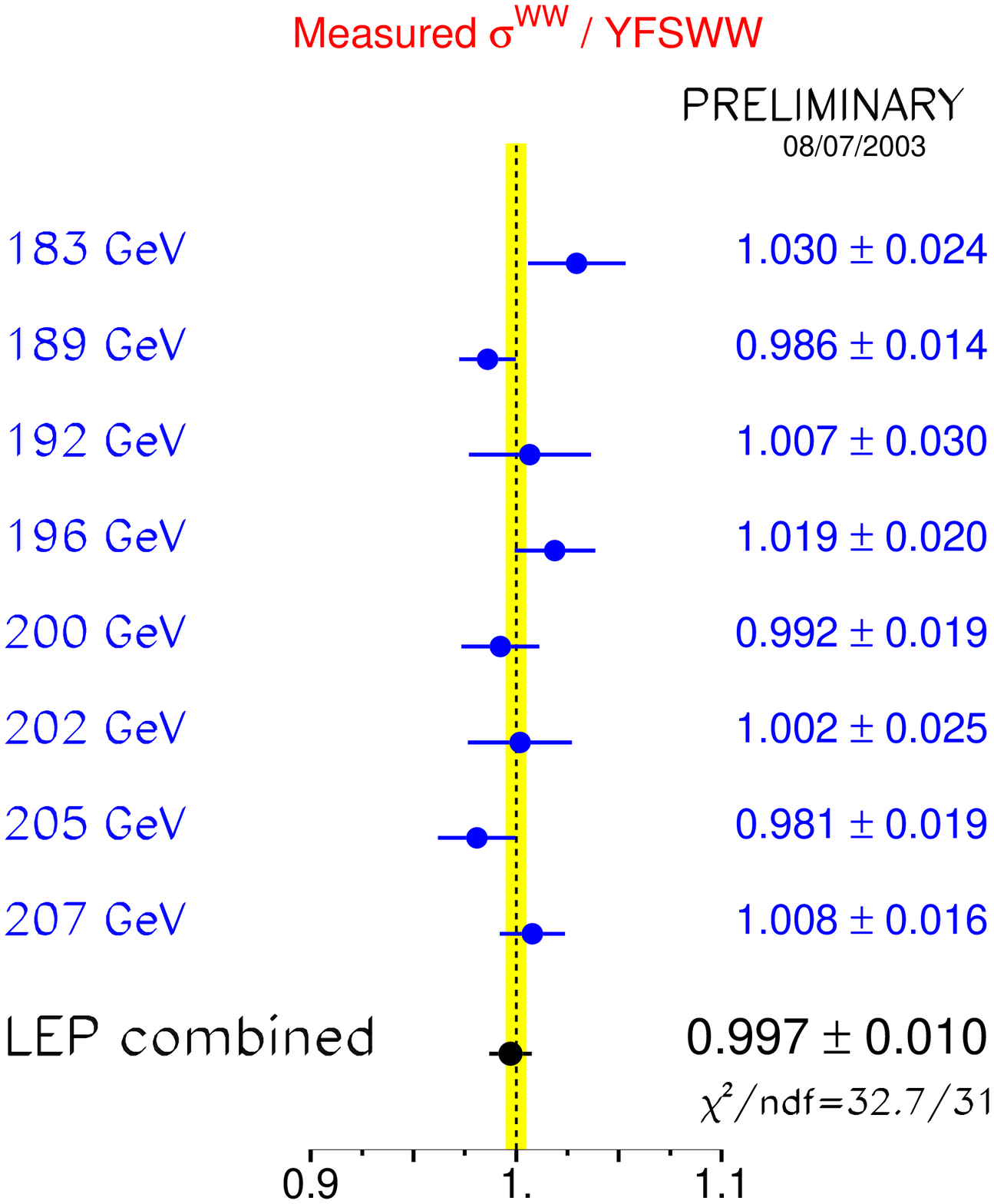,width=6.0cm}
\caption{  Ratio of the measured cross-section to the  prediction.
 \label{fig:w_crossb}}
}
\end{figure}

In Figures~\ref{fig:w_crossa} and~\ref{fig:w_crossb}  the measured  W-pair cross-section is compared
to predictions of RACOONWW~\cite{bib:racon} and YFSWW~\cite{bib:YFSWW}. 
The LEP measurements have reached a precision of ~1\% and agree
with the predictions.

\begin{figure}[!thb]
\parbox{5.6cm}{
\epsfig{file=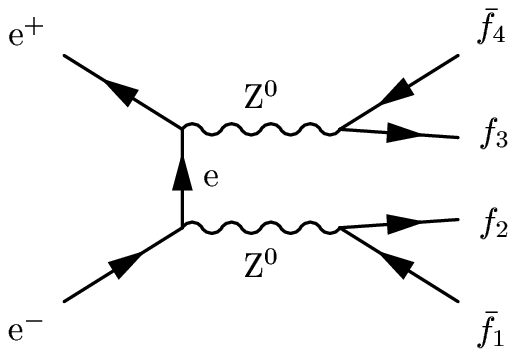,width=4.cm}
\epsfig{file=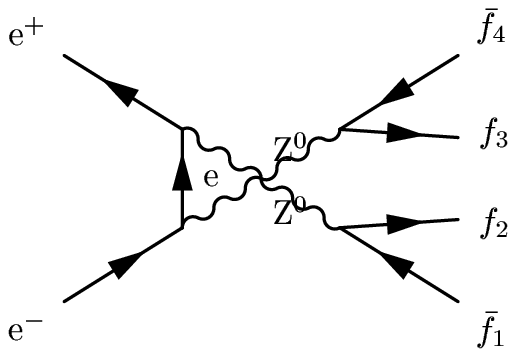,width=4.cm}
\caption{ Feynman diagrams for \Z pair production.
 \label{fig:z_crossa}}
}
\hspace*{.3cm}
\parbox{6.2cm}{
\epsfig{file=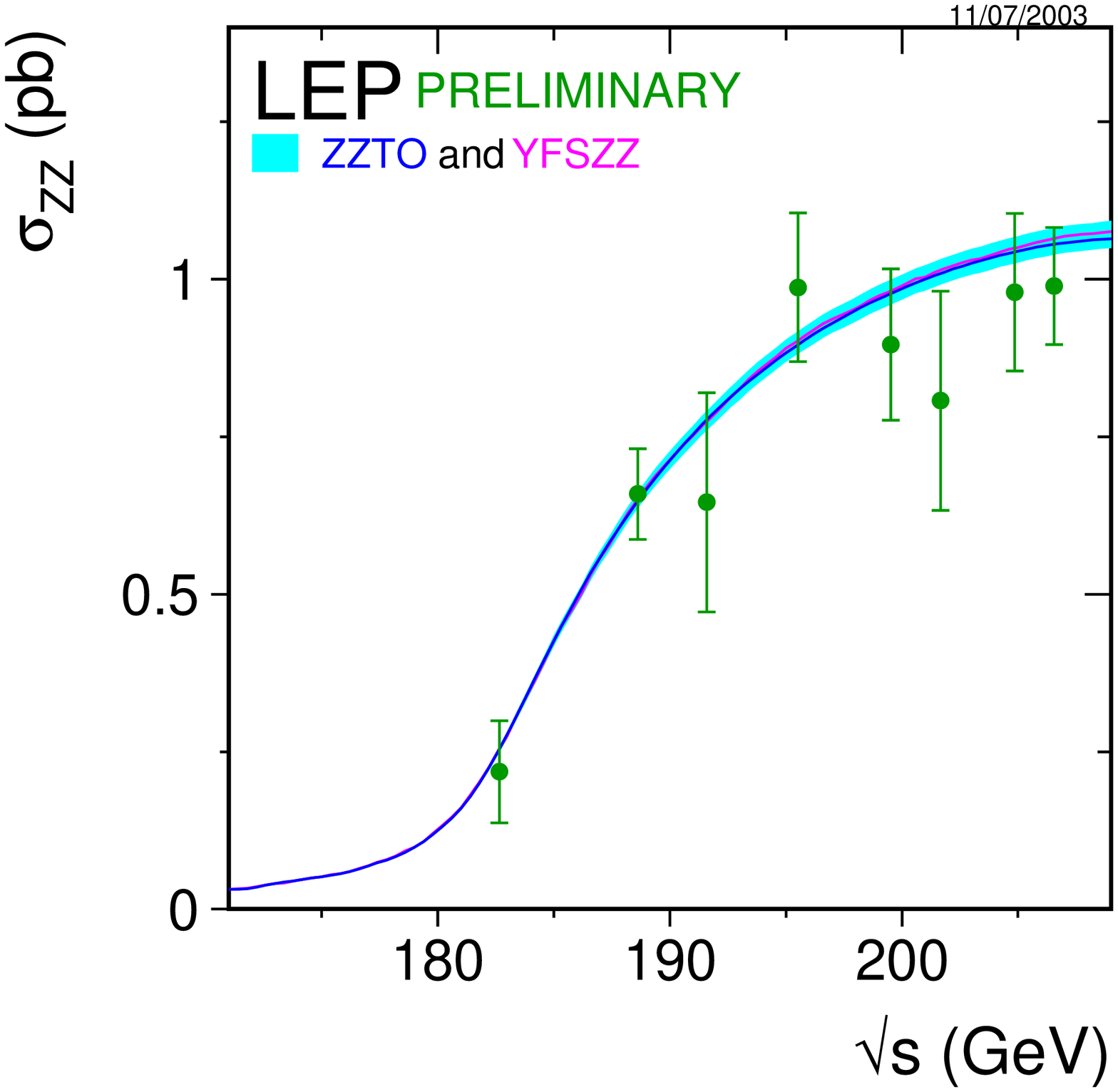,width=6.1cm}
\caption{ Z pair cross-section as function of center of mass energy.
 \label{fig:z_crossb}}
}
\end{figure}

Figure \ref{fig:z_crossa} shows the tree level Feynman diagrams contributing
to the Z-pair production (called NC02). 
Figure \ref{fig:z_crossb} shows that the  Z-pair cross-section agrees well
with  the theoretical predictions of ZZTO~\cite{bib:ZZTO} and YFSZZ~\cite{bib:YFSZZ}.
  
\section{W-mass determination}
In principle the invariant mass of two jets coming from the same
hadronic W  decay could be used to determine the most likely 
W-mass in a given event. This would however result in a poor mass
resolution of about 10\%. This resolution is dominated by the jet
energy resolution. 
The mass resolution can  be 
greatly improved by a kinematic fit in which the directions and
energies of the jets and the charged lepton 
(in the case of semileptonic W-pair events)
are allowed to vary within there uncertainties while requiring 
energy and momentum conservation.
Due to the constrain of energy conservation the relative uncertainty
in the determination of the beam energy will result in a corresponding
relative uncertainty on the W-boson mass.

The mass of the W-boson can be determined by comparing the reconstructed
mass distribution in data with Monte Carlo templates
corresponding to different W-boson masses and  
by minimizing the difference between  data and  Monte Carlo distribution. 
The  comparison can be extended to several dimensions using e.g. the
results from different kinematic fits or both the fitted mass and its uncertainty.

The optimal use of the information can be made by
using event probabilities. The event probabilities are calculated by the convolution
of a resolution function with a physics function. The physics function expresses the
probability to produce an event where the two produced W-bosons have masses 
$m_{\W_1}$ and  $m_{\W_2}$ for a given value of the W-boson mass. The physics function
is basically given by a Breit-Wigner function for the W decay modified by phase space
effects.
The resolution function parametrises the probability 
for the observation of a certain kinematic event topology, given that the 
produced  W-bosons have masses  $m_{\W_1}$ and  $m_{\W_2}$. The resolution function is in
the simplest case a Gaussian with the central value and width determined
by the kinematic fit. The mass of the W-boson is obtaind by maximising the total likelhood
which is given by the product of the event probabilties.
\section{Systematic errors}
\begin{table}
 \begin{center}
  \begin{tabular}{|l|r|r||r|}\hline
       Source  &  \multicolumn{3}{|c|}{Systematic Error on $M_\W$ ($\MeV$)}  \\  
                             &  $\qqln$ & $\qqqq$  & Combined  \\ \hline   
 Detector Systematics        & 14 &  10 & 14 \\
 Hadronisation               & 19 & 18 & 18 \\
 LEP Beam Energy             & 17 & 17 & 17 \\
 Colour Reconnection         & $-$& 90  & 9 \\
 Bose-Einstein Correlations  & $-$& 35 &   3 \\ \hline
\end{tabular}
 \caption{Major errors for the combined \LEP~W-mass results. }
 \label{tab:sys_com}
\end{center}
\end{table}
The major systematic errors are summarized in Table 1~\cite{bib:elw}.
For the W-mass measurement an excellent understanding of the detector response
is important. 
The energy scale and resolution, the 
angular resolution and their uncertainties 
have to be determined from the data. 
One important sample for the calibration of the detector are events which were
collected at the \Znull~resonance each year.
The jet or lepton pairs from the \Znull~decay are  back to back
and the total energy is equal to the beam energy.
In order to obtain information for jets and leptons with  energies different than half the
\Znull~mass, 3-jet events and events with an identified initial state photon were used
as well.

The hadronisation of coloured quarks and gluons into observable hadrons can only
be described with models. Despite energy and momentum conservation
in the hadronisation process this leads to a systematic uncertainty due to the following
effects: 
In the detector only particles with momenta larger than a given threshold are observed; 
the energy resolution for
neutral hadrons like $K_L$ and neutrons is quite low;  and for all charged particles
the pion mass is assumed in the calculation of the particle energy. 
The systematic uncertainty  due to  hadronisation is estimated by comparing different 
Monte Carlo Models (and sets of Monte Carlo parameters) which all describe
the high statistics $Z^0$ data well.  
It is important that, the  parameter sets 
 do not only describe inclusive distributions but also
reflect our  knowledge on exclusive rates like baryon and kaon fractions.

In the case that both W-bosons decay hadronically the uncertainty due to possible final 
state interactions between the decay products of the two W-bosons is by far the largest source
of systematic uncertainty. 
The possible bias on the reconstructed W-mass due to Bose-Einstein Correlations 
between identical bosons from the decay of different W-bosons and of colour reconnections 
between partons from different W-bosons can only be estimated with phenomenological
models.
Measurements of the correlations between identical bosons and measurements of
multiplicities, energy and particle flows (which are effected by possible colour
reconnections) are used to estimated the possible size of finale state interactions.
Thereby limiting the range of models and model parameters  used to estimate 
the systematic uncertainty on the W-boson mass.
The final state interactions predominately effect low momentum particles far
away from jets. The exclusion of this particles in the calculation of the jet direction
can greatly reduce the effect of final state interactions while the statistical
power of the mass determination is only slightly deteriorated.
This approach has not been yet used for the preliminary results shown is this presentation
but it will be used in the final publications of the W-mass measurements.

\section{Results}

\begin{figure}
\parbox{6.0cm}{
\epsfig{file=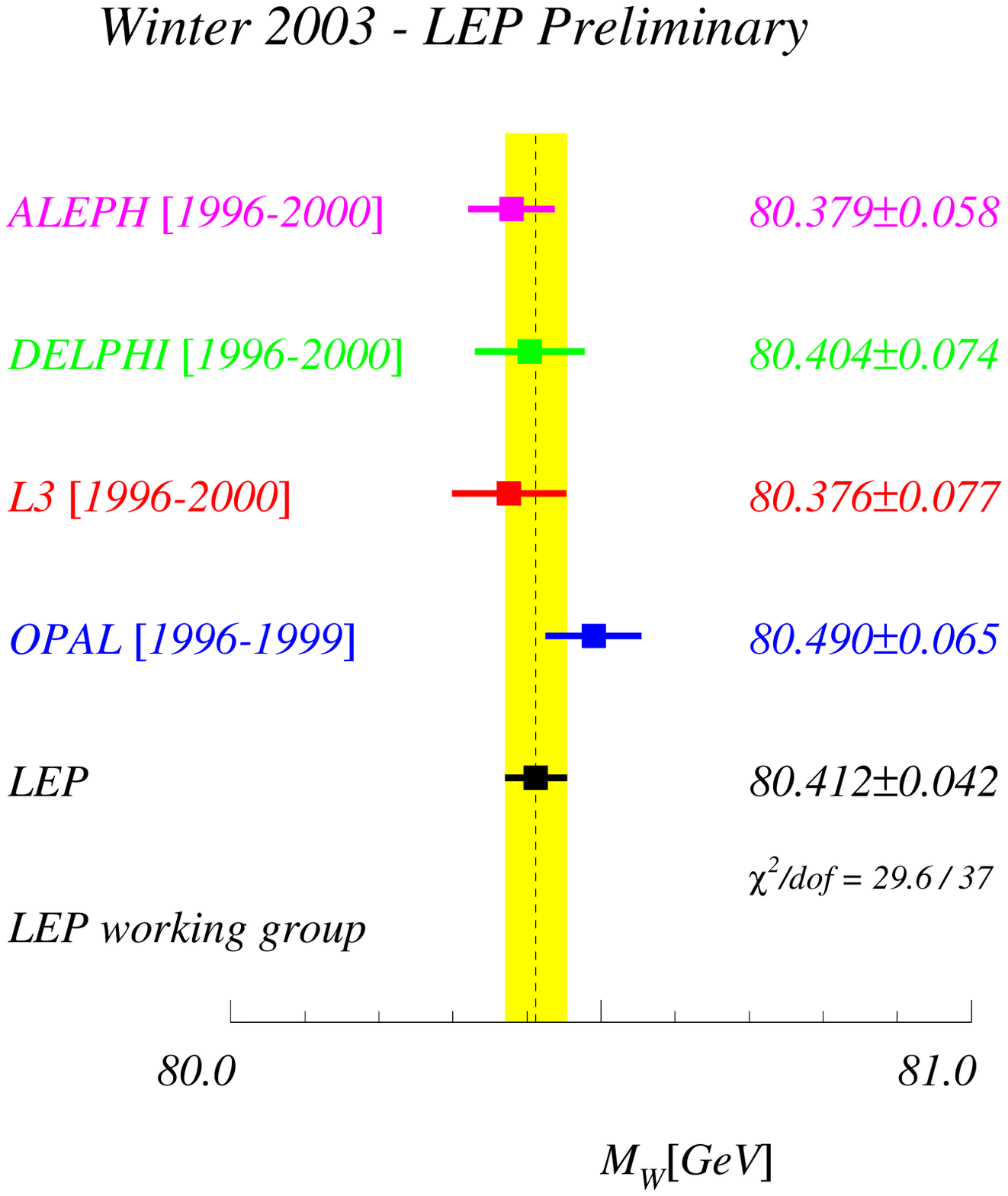,width=5.7cm}
\caption{Preliminary results for the measurement of the W-boson mass
  by the LEP collaborations 
 \label{fig:resa}}
}
\hspace*{.3cm}
\parbox{6.0cm}{
\epsfig{file=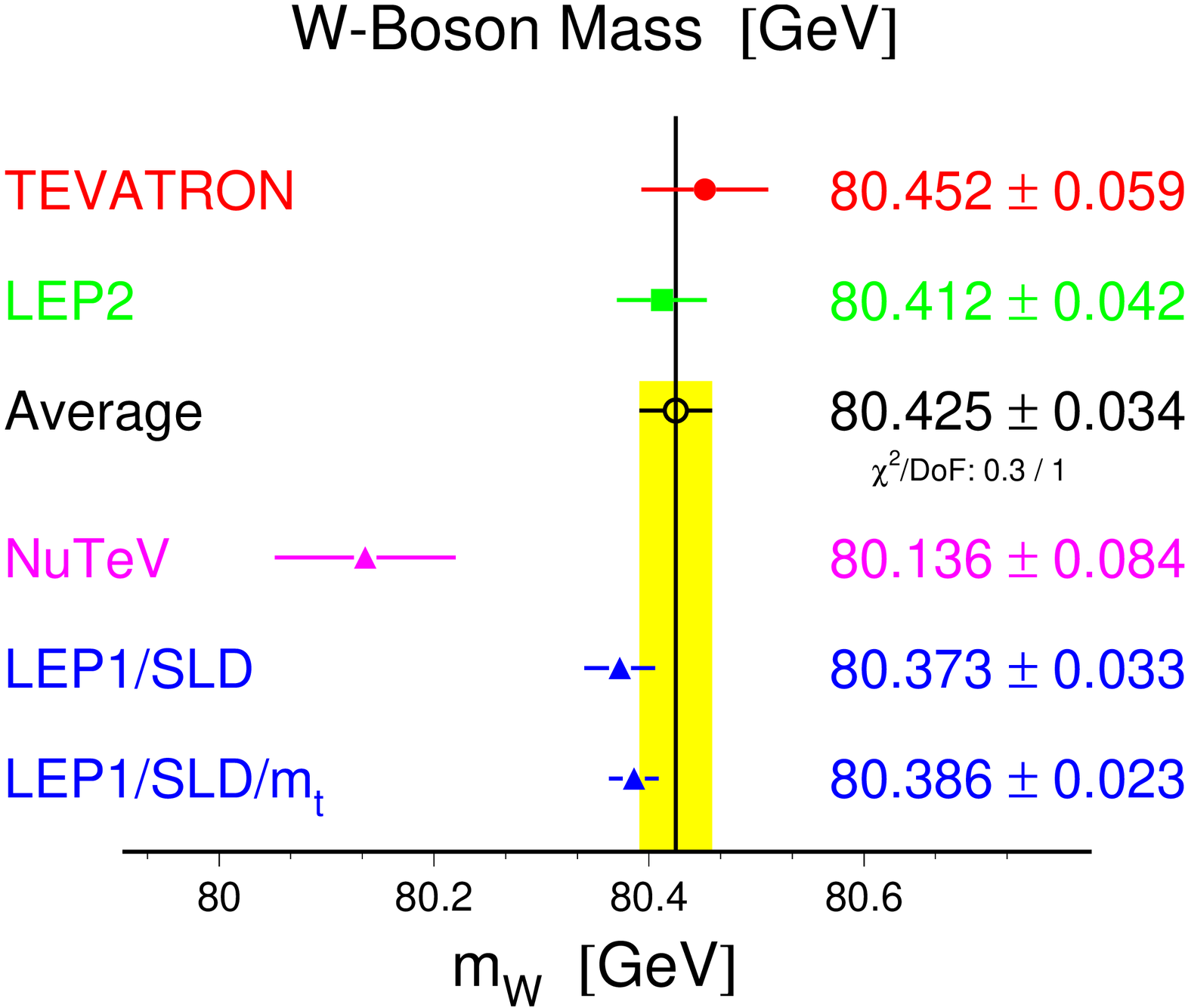,width=5.8cm}
\caption{Comparison of the direct measurements with
indirect predictions from electroweak precision measurements.
 \label{fig:resb}}
}
\end{figure}

Figure~\ref{fig:resa}  shows the preliminary results for the mass of the
W-boson from the four LEP collaborations. The combined result including the threshold measurements is~\cite{bib:elw}:

\[ M_\W = 80.412 \pm 0.042 \   \GeV. \]

The results of the direct measurements are compared in Figure~\ref{fig:resb} 
and Figure~\ref{fig:mw_mt} with indirect predictions from the fit to electroweak
precision measurements. One can see that the direct measurements have 
reached the same precision as the indirect predictions. Since no significant 
discrepancies are found the Standard Model predictions are confirmed at the
level of loop corrections.

\section*{Acknowledgements} 
I thank the LEP Electroweak WG for providing much
of the information presented here and the CERN accelerator devison for the
efficent operation of the LEP accelerator.


\begin{thebibliography}{0}

\bibitem{bib:rst-paper}R.~Str\"ohmer,
Int.\ J.\ Mod.\ Phys.\ A {\bf 18} (2003) 5127
\bibitem{bib:elw} 
The \LEP~Collaborations ALEPH, DELPHI, L3, OPAL, the \LEP~Electroweak Working Group
and the SLD Electroweak and Heavy Flavour Working Group. {\it``A Combination
of Prelimanry Electroweak Measurments and Constraints on the Standard Model''}, CERN-EP.2003-091.
\bibitem{bib:racon} A. Denner, S. Dittmaier, M. Roth and D. Wackeroth, \NPB{560}{1999}{33};
\NPB{587}{2000}{67}; \PRB{475}{2000}{127}.
\bibitem{bib:YFSWW} S. Jadach,W. Placek, M. Skrzypek, B.F.L. Ward, \PRD{54}{1996}{5434}. \\
 S. Jadach,W. Placek, M. Skrzypek, B.F.L. Ward, Z. W\c{a}s, \PLB{417}{1998}{326};
\PRD{61}{2000}{113010}; \CPC{140}{2001}{432}.
\bibitem{bib:ZZTO}  G. Passarino in
 {\it``Four Fermion Production in Electron Positron
Collisons"}, in {\it``Reports of the Working Groups on Precision Calculations for LEP2
Physics''}, CERN 2000-009, hep-ph-0005309.
\bibitem{bib:YFSZZ}  S. Jadach,W. Placek, B.F.L. Ward, \PRD{56}{1997}{6939}.


\end{thebibliography}
\end{document}